\documentclass[12pt]{article}
\usepackage{epsfig}
\usepackage{color}
\usepackage{amssymb,amsmath}

\newcommand{\bea}{\begin{eqnarray}}
\newcommand{\eea}{\end{eqnarray}}

 \makeatletter
\def\alt{\mathrel{\mathpalette\gl@align<}}
\def\agt{\mathrel{\mathpalette\gl@align>}}
\def\gl@align#1#2{\lower.6ex\vbox{\baselineskip\z@skip\lineskip\z@
\ialign{$\m@th#1\hfil##\hfil$\crcr#2\crcr\sim\crcr}}} \makeatother

\begin{document}
\begin{flushright}
\end{flushright}
\vspace*{1.0cm}

\begin{center}
\baselineskip 20pt 
{\Large\bf 
The PAMELA and Fermi signals \\
 from long-lived Kaluza-Klein dark matter 
}
\vspace{1cm}

{\large 
Nobuchika Okada and Toshifumi Yamada
} \vspace{.5cm}

{\baselineskip 20pt \it
Institute of Particle and Nuclear Studies, \\ 
High Energy Accelerator Research Organization (KEK)  \\
and \\
Department of Particles and Nuclear Physics, \\
The Graduate University for Advanced Studies (SOKENDAI), 
\\
1-1 Oho, Tsukuba, Ibaraki 305-0801, Japan} 

\vspace{.5cm}

\vspace{1.5cm} {\bf Abstract}
\end{center}

We propose a simple extension of the minimal 
 universal extra dimension model 
 by introducing a small curvature. 
The model is formulated as a small anti-de Sitter 
 curvature limit of the five-dimensional standard model (SM) 
 in the Randall-Sundrum background geometry. 
While the lightest Kaluza-Klein (KK) particle can be 
 thermal relic dark matter as usual in the universal extra dimension model, 
 the KK-parity is explicitly broken in the presence of 
 the small curvature and the KK dark matter decays 
 into the SM fermions with a long lifetime. 
Couplings of the KK dark matter with SM fermion pairs 
 in the five-dimensional bulk are controlled by 
 fermion bulk masses. 
By tuning bulk masses of quarks, we can suppress KK dark matter
 decay into quarks.
Further with a suitable choice of bulk masses for leptons, 
 KK dark matter decay into leptons 
 can account for the cosmic-ray electron/positron excesses 
 reported by the recent PAMELA and Fermi LAT satellite experiments.

\thispagestyle{empty}

\newpage

\addtocounter{page}{-1}
\setcounter{footnote}{0}
\baselineskip 18pt
%
Although the existence of (cold) dark matter has been established, 
 its identity remains a mystery in particle physics and cosmology. 
Pair annihilations or decays of the dark matter particles in the halo 
 associated with our Galaxy can produce cosmic rays 
 which could be detected by ongoing and future planned experiments. 
If observed, the properties of the dark matter can be 
 indirectly identified.

Recently, the PAMELA satellite experiment \cite{PAMELA} has reported 
 a significant cosmic-ray positron fraction 
 without a corresponding increase of the cosmic ray antiproton fraction. 
This results seems to be consistent with the previous results 
 of the HEAT \cite{HEAT} and AMS-01 \cite{AMS} experiments. 
In addition, the ATIC \cite{ATIC} and PPB-BETS \cite{BETS} 
 have reported an excess of the total positron plus electron flux 
 at energies around 100-800 GeV. 
Very recently, the Fermi LAT Collaboration has released 
 high quality data on the sum of electrons and positrons 
 from 20 GeV to 1 TeV \cite{Fermi}. 
Although the Fermi LAT data do not confirm the peak claimed by ATIC/PPB-BETS, 
 they still seem to indicate an excess compared to the expected background.

While these excesses could be reasonably explained 
 by some astrophysical source nearby \cite{aps}, 
 interpretation in terms of dark matter annihilations 
 or decays are particularly interesting 
 in the particle physics point of view. 
Since the PAMELA data has released, dark matter interpretations 
 of the excesses of the cosmic-ray fluxes have been 
 intensively investigated. 
One remarkable point is that the PAMELA data show 
 no excess for the cosmic ray antiproton fraction, 
 which implies a leptophilic nature of the dark matter. 
Several dark matter models possessing this nature 
 have been proposed \cite{models} 
 and dark matter pair annihilation or dark matter decay  
 dominantly into leptons can account for the excesses 
 of the cosmic-ray fluxes. 
After the very recent Fermi LAT data, it has been argued 
 \cite{GRASSO} (see, also, \cite{BEZ}) that 
 models where the dark matter only pair annihilates into charged leptons 
 can give a satisfactory fit to both the PAMELA and Fermi LAT data 
 for the dark matter masses between 400 GeV and 2 TeV 
 with a suitable annihilation cross section, 
 $ \langle \sigma v \rangle =10^{-24}-10^{-23}$ cm$^3$/s. 
For the scenario of a long-lived dark matter particle decaying 
 only into charged leptons, this cross section is translated 
 to the lifetime of the dark matter through the relation 
 $\rho^2 \langle \sigma v \rangle /(2 M^2) \leftrightarrow 
 \rho/(M \tau) $ with a local dark matter density 
 $\rho \sim 0.3$ GeV/cm$^3$, 
 and we find the dark matter with a mass 
 in the range of 800 GeV-4 TeV and a lifetime $\tau \sim 10^{27}$ sec 
 can fit the data.

Here we note that both scenarios of the dark matter 
 pair annihilation and the long-lived dark matter 
 have some phenomenological issues. 
In the scenario of the dark matter pair annihilation, 
 the pair annihilation cross section consistent with 
 the dark matter observed relic abundance is too small 
 to fit the cosmic-ray excesses, so that the ``boost'' factor 
 as big as 100-1000 is necessary to enhance 
 the annihilation cross section of dark matter in the halo. 
This boost factor could either have an astrophysical origin 
 (large inhomogeneity of dark matter distribution) 
 or have a particle physics origin such as 
 the Sommerfeld enhancement \cite{Sommerfeld} 
 and Breit-Wigner enhancement \cite{B-W}. 
However, according to the result of recent $N$-body simulations \cite{N-body}, 
 the probability of forming such a large inhomogeneity may be quite small. 
On the other hand, in order to accommodate the particle physics 
 mechanism for enhancing the cross section, 
 we need to elaborate models for dark matter. 
In the scenario of a long-lived dark matter 
 (gravitino as the lightest superpartner in supersymmetric 
  models with $R$-parity  violation is a typical example), 
 the dark matter superweakly couples with the standard model(SM) particles 
 and cannot be in thermal equilibrium. 
Thus, the relic density of the dark matter highly depends 
 on the history of the early Universe such as reheating 
 temperature after inflation. 
In this paper, we propose a model for dark matter 
 which can account for the cosmic-ray electron/positron excesses 
 without these issues.

There have been many models proposed which can naturally 
 provide a dark matter particle and among them, we consider 
 the minimal universal extra dimension(UED) model \cite{UED1, UED2} in this paper. 
In this model, the 1st Kaluza-Klein(KK) mode of the SM U(1)$_Y$ gauge boson 
 is the lightest KK particle (LKP) and the candidate for the dark 
 matter thanks to the conservation of KK parity \cite{UED3}. 
Thermal relic abundance of the KK dark matter was first 
 investigated in Ref.~\cite{LKP1}. 
Through more elaborate analysis by taking into account 
 all possible processes such as coannihilations with 
 other KK particles and higher KK particles resonances \cite{LKP2}, 
 the allowed region of the KK dark matter mass has been found 
 to be in the range of 600 GeV-1.3 TeV \cite{KMS}. 
A KK dark matter mass $\gtrsim 800$ GeV may be favorable 
 otherwise the KK graviton can be the LKP. 
The region $\gtrsim 1.3$ TeV is excluded, 
 since in this case, the 1st KK mode of the charged Higgs boson  
 becomes the LKP \cite{KMS}.

We extend the minimal UED model by introducing a small curvature. 
In fact, we define the model as the limit of a very small 
 anti de Sitter(AdS) curvature of the five-dimensional SM 
 \cite{bulkSM1, bulkSM2, bulkSM3} 
 in the Randall-Sundrum (RS) warped background geometry \cite{RS}. 
In the presence of the curvature, the KK parity is explicitly broken 
 and hence the KK dark matter is no longer stable and, in fact, 
 decays into SM fermion pairs. 
We tune an extremely small curvature so as to make the dark matter 
 long-lived with lifetime $\tau \sim 10^{27}$ sec. 
Such a small curvature has no effect on the relic abundance 
 of the dark matter, while the dark matter decay can produce 
 the cosmic rays. 
The branching ratio of the dark matter decay depends on the mass parameters 
 of bulk fermions and hence controllable. 
It is even possible to completely eliminate a specific decay mode 
 by tuning bulk fermion masses. 
By suitably fixing free parameters in our model, 
 we can provide a set of parameters chosen to give 
 a satisfactory fit to both the PAMELA and Fermi LAT data 
 in Ref.~\cite{GRASSO}, while satisfying the correct thermal 
 relic abundance of the KK dark matter with mass 800 GeV-1.3 TeV.

Let us now formulate our model. 
We consider the 5D SM model in the RS warped background geometry, 
 in which the fifth dimension is compactified on the orbifold $S^1/Z_2$ 
 with radius $r_c$. 
The 5D metric is given by \cite{RS}  
\bea 
 d s^2 = e^{-2 \sigma(y)} \eta_{\mu \nu} dx^{\mu} dx^{\nu}  - dy^2, 
\eea 
 where $ \sigma(y) = k \vert y \vert$ 
 for $-\pi r_c \leq y \leq \pi r_c $ with the AdS curvature $k$. 
%

The action for a bulk fermion in the RS background geometry 
 is given by \cite{bulkSM1, bulkSM2, bulkSM3} 
\bea 
 S_f = \int d^4x \int dy e^{-4 \sigma} 
   \left[ V_n^M 
   \left( \frac{i}{2} \overline{\Psi} \gamma^n {\cal D}_M \Psi + h.c. \right)
   -{\rm sgn}(y) (c k) \overline{\Psi} \Psi
   \right], 
\eea
 where $n,M=0,1,\cdots, 4$, 
 ${\cal D}_M = \partial_M - i g_5 A_M$ 
 is the 5D covariant derivative with a 5D gauge field $A_M$, 
 $V_\mu^M =e^\sigma \delta_\mu^M$, $V^4_4=1$, 
 and $\gamma^n=(\gamma^\mu, i \gamma_5)$. 
The bulk fermion mass term is defined as $c k$ 
 with a dimensionless constant $c$. 
We expand the bulk fermion by the KK modes as 
\bea
 \Psi (x,y) = 
   \sum_{n=0}^{\infty} \psi^{(n)}_{L}(x) e^{2 \sigma(y)}f^{(n)}_{L}(y) 
 + \sum_{n=0}^{\infty} \psi^{(n)}_{R}(x) e^{2 \sigma(y)}f^{(n)}_{R}(y).  
\eea
Because of the requirement of the $Z_2$ symmetry of the action, 
 $f^{(n)}_{L}(y)$ and $f^{(n)}_{R}(y)$ must have opposite 
 $Z_2$ parity. 
In this paper, we treat all the SM fermions as left-handed 
 and choose $f^{(n)}_{L}(y)$ to be $Z_2$-even, 
 so that zero-mode left-handed fermions are identified as the SM fermions. 
Solving the equation of motion for KK modes 
 with suitable boundary conditions at the orbifold fixed points 
 corresponding to the $Z_2$-parity assignment, 
 we have zero-mode wave functions such as 
\bea 
 f^{(0)}_{L}(y)= N^{0}_{c} e^{c \sigma(y)} , ~~~ f^{(0)}_{R}(y)= 0 
\eea 
 with a normalization factor 
\bea 
 N^{0}_{c}  = \sqrt{ 
   \frac{ k (1+2c)}{2 \left(  e^{k \pi r_{c} (1+2c)} - 1  \right)}} . 
\eea

Next, we consider a 5D gauge field ${\cal B}_M(x,y)$ 
 and expand its wave function as 
\bea 
 {\cal B}_{\mu}(x,y) = \sum_{n=0}^{\infty} \ B^{(n)}_{\mu}(x) \ f^{(n)}_{B}(y)
\eea 
 with the gauge choice ${\cal B}_4=0$. 
The $Z_2$-parity of the wave functions is chosen to be even. 
Solving the equation of motion for the KK modes with suitable 
 boundary conditions \cite{bulkSM1}, the wave functions for zero mode 
 and the 1st KK mode are found to be 
\bea 
 f^{(0)}_{B}(y) &=& \sqrt{ \frac{1}{ 2\pi r_{c} } } , \nonumber \\ 
 f^{(1)}_{B}(y) &=& N^{1}_{B} \ e^{\sigma(y)} 
 \left[ 
   J_{1} \left( \frac{m_{1}}{k} e^{\sigma(y)} \right) 
   - \frac{J_{0} (\frac{m_{1}}{k}) }{Y_{0} (\frac{m_{1}}{k}) } 
   Y_{1} \left( \frac{m_{1}}{k} e^{\sigma(y)} \right) \right]  ,
\end{eqnarray}
where 
 $J_q$ and $Y_q$ denote Bessel functions of order $q$, 
 the normalization factor $N_B^1$ is given by 
\bea 
  1/N^{1}_{B} = 
  \sqrt{ 
   \int^{\pi r_c}_{-\pi r_c} dy 
   e^{2 \sigma(y)} \left[ 
    J_1 \left( \frac{m_1}{k} e^{\sigma(y)} \right) 
     -  \frac{J_{0} (\frac{m_{1}}{k}) }{Y_{0} (\frac{m_{1}}{k}) } 
    Y_1 \left( \frac{m_1}{k} e^{\sigma(y)} \right) \right]^2 },  
\eea
 and the 1st KK mode mass is obtained as 
 $m_1 = z_1 k $ with $z_1$ being the first positive solution of  
\bea
  J_{0} ( z_1 e^{\pi k r_{c}} )  Y_{0} (z_1)  - 
  J_{0} (z_1) Y_{0} ( z_1 e^{\pi k r_{c}} ) =  0 . 
\label{eigenvalue}
\eea

Now we identify the bulk gauge field as the SM U(1)$_Y$ gauge field 
 and the U(1)$_Y$ gauge coupling in 4D is defined as 
\bea 
  g_Y = g_{5} \int_{-\pi r_{c}}^{\pi r_{c}} dy 
     e^{\sigma(y)}  f^{(0)}_{B}(y) \ f^{(0)}_{L}(y) \ f^{(0)}_{L}(y) 
     = \frac{g_5}{\sqrt{2 \pi r_c}} 
\eea 
 with the 5D gauge coupling $g_5$, 
 which is independent of the bulk fermion mass term $c$. 
In the same way, the effective gauge coupling 
 among the 1st KK gauge boson and a pair of SM fermions 
 is given by 
\bea 
  g_Y^{(1)}(c) = g_{5} \int_{-\pi r_{c}}^{\pi r_{c}} dy 
     e^{\sigma(y)}  f^{(1)}_{B}(y) \ f^{(0)}_{L}(y) \ f^{(0)}_{L}(y),  
\eea 
 which depends on the bulk mass parameter $c$.

We now consider our model which is a simple extension of 
 the minimal UED model with a small curvature, 
 and we take $1/r_c \sim 1$ TeV for a typical KK particle 
 mass scale in the minimal UED model and 
 $\sigma_\pi =\pi k r_c \ll 1 $ to obtain almost flat extra dimensions. 
In this case, $z_1 \gg 1$ in Eq.~(\ref{eigenvalue}). 
Using the asymptotic expansion of Bessel functions, 
\bea 
 J_{\nu}(z) & \sim & \sqrt{ \frac{2}{\pi z}} 
                     \cos( z - \frac{2 \nu + 1}{4} \pi ) 
 \nonumber \\ 
 Y_{\nu}(z) & \sim & \sqrt{ \frac{2}{\pi z}} 
                     \sin( z - \frac{2 \nu + 1}{4} \pi ) 
\eea 
 for $z \gg 1$ and $e^{\pi k r_c} \sim 1 + \sigma_\pi$, 
 Eq.~(\ref{eigenvalue}) is reduced into $ \sin (z_1 \sigma_\pi)=0$
 and the 1st KK mode mass is approximately given as 
 $m_1 \simeq 1/r_c $, which is consistent with the 1st KK particle mass 
 at the tree level in the UED model.

Let us evaluate the effective coupling among the 1st KK mode 
 of the SM U(1)$_Y$ gauge boson (KK dark matter) 
 and a pair of SM fermions with the bulk mass $c k$. 
At the leading order of the small parameter $\sigma_\pi$, 
 we find  
\bea 
 g_Y^{(1)}(c) 
  &=& g_{5} \int_{-\pi r_{c}}^{\pi r_{c}} dy 
     e^{\sigma(y)}  f^{(1)}_{B}(y) \ f^{(0)}_{L}(y) \ f^{(0)}_{L}(y)
\nonumber \\
  &\simeq& g_{Y} \sigma_\pi \frac{1+2c}{e^{(1+2c) \sigma_\pi}-1}
  \frac{\int^{1}_{0} d t e^{ \left( \frac{3}{2}+2c \right) \sigma_\pi t} \cos(\pi t) }
       {\int^{1}_{0} d t e^{ \sigma_\pi t} \cos^{2}(\pi t) } 
\nonumber \\
  &\simeq& g_{Y} \left[- \frac{ \sqrt{2} }{\pi^{2}} (3 + 4 c) \right] 
   \sigma_\pi . 
\end{eqnarray}
The effective coupling is controlled by the bulk mass parameter $c$ 
 and becomes zero in the flat space-time limit (UED limit )  
 $\sigma_\pi \to 0$, as expected. 
Note that we can obtain a vanishing coupling 
 (in the leading order of $\sigma_\pi$) for $c=-3/4$.

In our model, all bulk mass terms for SM chiral fermions  
 are free parameters and depending on the bulk masses, 
 we can consider a variety of decay modes of the KK dark matter. 
For example, the leptophilic nature of the dark matter decay 
 can be obtained by fixing $c$'s for bulk quarks to be around $-3/4$. 
The flavor dependence of the KK dark matter decay mode 
 corresponds to the fermion bulk masses.

The partial decay width of the dark matter into left-handed fermion
 pairs is given by
\bea 
 \Gamma ( B_{\mu}^{(1)}  \rightarrow \psi^{i}_{L} \bar{\psi}^{i}_{L}) 
  = N_i Y_i^{2} g^{(1)}_{Y}(c_i)^{2} \frac{m_{1}}{24 \pi},  
\eea 
 where we have neglected the fermion mass, 
  $i$ denotes flavor, 
  $Y_i$ is the corresponding hypercharge, 
  $N_i$ denotes the number of degrees of freedom 
  of final state fermions, for example, 
  $ N_{i} = 3$ for SU(2) singlet quarks. 
The lifetime of the KK dark matter is 
\bea
 \tau \simeq 2 \times 10^{28} {\rm sec} \times 
 \left(\frac{10^{-26}}{\sigma_\pi}\right)^2  
 \left(\frac{1 {\rm TeV}}{m_1} \right) 
 \left[ \sum_{i} N_i Y_i^2 \left( 1 + \frac{4}{3} c_i \right)^2 
  \right]^{-1} . 
\eea 
In order to give $\tau \sim 10^{27}$ sec 
 suitable for explanation of the PAMELA and Fermi LAT data, 
 we need to tune the AdS curvature to be an extremely small value: 
\bea 
 k \sim 10^{-15} {\rm eV} \times \left(\frac{m_1}{1 {\rm TeV}}\right). 
\eea 
We may think of this fine-tuning problem as a cosmological constant 
 problem, because the AdS curvature in the RS model originates 
 from the (negative) bulk cosmological constant.

So far we have only considered the decay modes 
 of the KK dark matter to the SM fermion pairs. 
In precise, the KK dark matter is a mixture of 
 1st KK modes of the U(1)$_Y$ and charge neutral SU(2) 
 gauge bosons after the electroweak symmetry breaking. 
Thus, the KK dark matter can decay into $W$-boson pairs 
 through the KK component of the charged neutral SU(2) 
 gauge boson. 
However, this decay amplitude is suppressed by 
 a small mixing angle $(m_Z/m_1)^2 \sim 0.01$ 
 with $Z$-boson mass $m_Z$ and $m_1 \sim 1$ TeV,  
 and is negligible compared to the decay amplitude 
 to SM fermion pairs.

In summary, we have proposed a simple extension of 
 the minimal UED model to account for the cosmic-ray 
 electron/positron excesses reported by the recent PAMELA 
 and Fermi LAT satellite experiments. 
We have formulated our model as the 5D SM in the RS background 
 geometry with a small AdS curvature. 
Since the curvature we have introduced is extremely small, 
 the LKP is thermal relic dark matter as usual in the minimal UED 
 model and the correct thermal relic abundance is realized 
 with a suitable mass range 600 GeV $\lesssim m_1 \lesssim$ 1.3 TeV. 
However, the KK parity is explicitly broken in the presence of 
 the AdS curvature and the KK dark matter is no longer stable. 
We have investigated the decay modes of the KK dark matter 
 into SM fermion pairs and found that the (partial) decay width 
 is controlled by the bulk fermion masses which are all 
 free parameters of the model. 
The special choice of the bulk mass term $c=-3/4$ 
 can even eliminate the coupling between the KK dark matter 
 and fermions in the leading order of the small AdS curvature. 
Therefore, the leptophilic nature of the KK dark matter, 
 which is implied by the PAMELA data 
 for the cosmic ray antiproton fraction, 
 can be achieved by fixing bulk masses for quarks around $-3/4$. 
The model can provide a suitable parameter set, 
 for example, $m_1= 800$ GeV-1.3 TeV and $\tau \sim 10^{27}$ sec,  
 which gives not only the correct thermal relic abundance of 
 the KK dark matter but also a satisfactory fit to 
 both the PAMELA and Fermi LAT data through the leptophilic decay 
 of the KK dark matter. 
In our model, the flavor structure of primary cosmic rays from 
 the KK dark matter decays corresponds to the bulk fermion mass terms. 
Precise measurements for a variety of cosmic-ray fluxes 
 in future experiments may fix some of SM fermion bulk masses \cite{OY}.

\section*{Acknowledgments}
This work of N.O. is supported in part by the Grant-in-Aid 
 for Scientific Research from the Ministry of Education, 
 Science and Culture of Japan, No.~18740170. 



\end{document}